# System Architecture of HatterHealthConnect: An Integration of Body Sensor Networks and Social Networks to Improve Health Awareness

Hala ElAarag, David Bauschlicher and Steven Bauschlicher
Department of Mathematics and Computer Science
Stetson University
421 N. Woodland Blvd.
DeLand, FL 32723

{helaarag, dbausch, sbausch} @stetson.edu

*ABSTRACT:*

*Over the last decade, the demand for efficient healthcare monitoring has increased and forced the health and wellness industry to embrace modern technological advances. Body Sensor Networks, or BSNs, can remotely collect users data and upload vital statistics to servers over the Internet. Advances in wireless technologies such as cellular devices and Bluetooth increase the mobility users experience while wearing a body sensor network. When connected by the proper framework, BSNs can efficiently monitor and record data while minimizing the energy expenditure of nodes in the BSN. Social networking sites play a large role in the aggregation and sharing of data between many users. Connecting a BSN to a social network creates the unique ability to share health related data with other users through social interaction. In this research, we present an integration of BSNs and social networks to establish a community promoting well being and great social awareness. We present the system architecture; both hardware and software, of a prototype implementation using Zephyr HxM heart monitor, Intel-Shimmer EMG senor and a Samsung Captivate smart phone. We provide implementation details for the design on the base station, the database server and the Facebook application. We illustrate how the Android application was designed with both functionality and user perspective in mind that resulted in an easy to use system. This prototype can be used in multiple health related applications based on the type of sensors used.*

## 1.INTRODUCTION

The health care industry has been rapidly expanding over the past few years. A particular area of research experiencing rapid discovery is the use of body sensor networks, or BSNs, to monitor patients. A BSN consists of sensors recording biological data which is then sent to the corresponding data coordinator. From there the data can be interpreted in various ways. At first, hospitals used BSNs to monitor patients on site, but soon technological advances allowed the patients to move into more native environments, such as their own homes, to be monitored. A wired connection between the sensors and data coordinator was originally used to facilitate BSN communication. With the advent of wireless protocols however, technologies such as Bluetooth and ZigBee (802.15.4) have eliminated the wires from the network and increased the mobility of the patient. Users can now wear body sensors and perform everyday tasks including exercising without the need to adjust their body sensors.

Mobile devices such as mobile phones and PDAs have seen significant development in the recent years as well. Modern day smart phones run complex operating systems such as Android and are more comparable to laptops than previous cell phones. Because of their inherit portability, these smart phones make an excellent candidate for the data controller portion of a BSN. These phones also come equipped with multiple sensors such as GPS, accelerometer, and light sensors, which further promote data aggregation and can reduce the need to buy additional sensors. Google's Android operating system is an excellent candidate for a BSN controller. The open source Android OS runs Java source code, and because of its portability and Bluetooth API, a single native controller can be made for the OS which could then be propagated to other devices in the future.

The final processing layer of BSN data can be seen as the most crucial part of the entire process. The raw data by itself is worthless unless it can be used to solve a problem or enhance the well being of the patient. The data can be used to track health patterns in patients, display vital health statistics, or even create recommendations for exercise.

This paper presents a system that submits sensor data into the realm of social networking. Social networking web sites such as Twitter and Facebook have gained massive popularity among users worldwide. User's send updates about their events, location, status, etc, only to share it with other users around the globe. If applied to the health industry, information such as health condition, weight, exercise levels, and heart rate, could be shared between users to create a positive and educational environment about wellness. The framework of this research has appeared in [3]. The rest of the paper is organized as follows: section 2 provides background information about body sensor networks, while section 3 presents related work in the literature. Section 4 gives a quick overview of the HatterHealthConnect system. Section 5 and 6 describe the hardware and software architecture of HHC, respectively, and section 7 discusses the actual implementation of the system. Finally, section 8 provides a conclusion of the paper while section 9 suggests some future work.

## 2. BACKGROUND

A body sensor network, or BSN, can almost be seen as a subset or derivative of a Wireless Sensor Network (WSN). Many of the same components and issues of WSNs are present in BSNs, and the similar techniques can be used to solve the problems. However, there are some distinctions between the two. BSNs' nodes consist of sensors able to read biological data, and these results often are forwarded to a lab or hospital to perform some medical evaluation. Sensor nodes can also consist of implantable devices. These implanted devices require a slightly different configuration because they only have one battery charge. Fixing them while implanted is currently not an option [1]. BSNs are often involved in critical systems as well; a patient's life or death may rely on the speed of the message delivery. Therefore, message delivery rate and congestion are important issues a BSN must focus on [2].

Most modern BSNs contain a variety of medical sensors along with a controller or base station. Many medical sensors such as a pulsioximeter, ECG, and heart rate monitor can be used to report data to the controller [7]. In order to work with an array of devices and reduce energy costs, most communicate through common protocols such as Bluetooth or ZigBee. The controller node must support the protocol as well, and often acts as an internet gateway too. Modern day cell phones are a natural fit for the controller node because they meet the protocol requirements and are portable [7]. Cell phones have additional connectivity tools such as SMS and 3G to increase network resources. Some BSNs rely solely on these technologies to communicate, and have even developed entire protocols around them [21]. The patient also has prior experience with his/her mobile device, so it is often an easy transition. Previous experiments have used similar technologies such as PDAs for the controller node [8].

Body sensor networks can greatly vary in size and purpose. A small BSN monitoring an athlete's vitals while exercising may only require a single patient. On the other hand, a large healthcare facility may have hundreds of patients, each with his/her sensors reporting to a large central database or program [10]. Depending on the case, a different network structure may be needed to maximize the efficiency of the BSN. However, the same basic BSN issues need to be addressed regardless of purpose or size. Energy efficiency is crucial for BSNs taking advantage of wireless sensors and devices, and can be improved by enhancing the network layers. The second issue BSNs face is security. Health data is extremely personal, and it is important for patients to know their data is secure and only accessible to the appropriate parties.

## 3. RELATED WORK

In this emerging field of research, other similar projects have been created. Harvard's CodeBlue [16] is a project created to adapt wireless sensor networks for use in emergency medical situations. In this project, sensor networks are used to send real-time vital signs from a group of patients to emergency medical technicians. These sensor networks would allow for a rapid medical response to a mass casualty event and would allow health personal to locate those patients in most need. CodeBlue makes use of the Berkeley MICA mote which contains a microcontroller, local storage area, and a low-power radio. These motes run on the TinyOS operating system and have a battery life of approximately 5-6 days while running. The communication is handled over the IEEE 802.15.4 standard to also conserve power. A pulse oximeter that attaches to the mote has been developed to deliver heart rate and blood-oxygen level, and an ECG was in development during the publishing of the paper. CodeBlue is highly scalable and works in an ad hoc setting, a critical need for use in emergency situations.

A WSN has been proposed by members of the Computer Science Department at the University of Virginia [17] to provide support to the increasing elderly population. This WSN combines wearable sensors along with environmental sensors to provide medical monitoring and memory enhancement to the patient. The architecture is divided into five main components: a body network, an emplaced sensor network, a backbone, back-end databases, and human interfaces. The body network is composed only of unobtrusive wireless sensors, but the environmental sensors have the option of being completely wireless and requiring batteries for power or of being plugged into an outlet. Data gathered from the body and emplaced sensor network is sent across the backbone of the WSN to be stored in the back-end databases or displayed in one of the human interfaces. The sensors used in this implementation include a motion sensor, temperature sensor, breathing rate sensor, pulse-oximeter, and an EKG. These sensors communicate with the backbone using the Zigbee (802.15.4) wireless protocol. Real-time data can be viewed by a PDA that can connect to the backbone or by LCDs located on the motes.

Another wireless sensor network, discussed in [18], is designed to safely and continuously send physiological data from a patient to a local WLAN to be further transmitted. The patient wears an ECG sensor which wirelessly transmits data to one of the local relay nodes that are strategically placed throughout the entire building the patient is living in. This overcomes the deadspot problem that could occur if using a WLAN through walls. In order to avoid excess power consumption, the ECG sensor periodically takes samples, stores them in a buffer, and then goes back to sleep. Only once the specified buffer is full does it place the frame around the information and send it to the local node. The nodes have two different modes, self organizing mode and data transmission mode. When a node initially starts up, it stays in the self organizing mode until the surrounding nodes successfully add the node into their routing table and vice versa. Once the node is added to the system, it waits for data and then acknowledges the received packets and forwards the data towards the uplink node. An SMS message would be sent in the case of an emergency found in the ECG data.

DexterNet [4], a heterogeneous body sensor network, is an open-source project that makes use of the open-source library called Signal Processing In Node Environment (SPINE). DexterNet has a three layer architecture that includes a body sensor layer (BSL), a personal network layer (PNL), and global network layer (GNL). The BSL contains two different types of custom sensors. The first is a motion sensor that contains an accelerometer and a gyroscope. The second type is a biological sensor that acts as an electrical impedance pneumography (EIP), and ECG, and an accelerometer. The PNL consists of a Nokia N800 tablet that communicates and collects data from the different sensors within the BSL. This communication is accomplished over IEEE 802.15.4 and makes use of the SPINE API on the node and base station side. Finally, the Nokia N800 forms a GNL by sending data across the Internet through a Bluetooth, Wifi, or other broadband connection. These Internet servers then use the data collected for higher level applications. Some server-side applications already created provide features such as displaying a graphical animation of a user's current position, creating a database of movement information to improve human movement recognition algorithms, and monitoring pollution on a patient's walk through the city [4].

Researchers at the University of California, Berkeley have begun working on an API for the Android OS that provides specific functionality for BSN development. The project, titled WAVE [5], has several core components including sensor interaction, database interaction, and data processing functions. The sensor interaction is handled by the SPINE framework, allowing WAVE to focus less on the low level communication between sensor nodes and the mobile platform. Because the Android OS has a java based development kit, SPINE should work well with it as long as the sensors can also make use of it. Also, because the Android OS is becoming increasingly popular on multiple models of phones, the SPINE framework can be ported over once and used for multiple applications on multiple devices. After data is collected from the nodes in WAVE, database interaction is handled using REST architecture so a user can easily insert and store information from the mobile platform into a remote database. This REST architecture basically refers to a stateless, client-server model similar to how the World Wide Web works. Finally, the data processing functions allow the user to access algorithms that are frequently used within BSN applications. These functions include action recognition, energy expenditure calculating, and GPS location tracking.

Using the WAVE API, a few applications have already been created to monitor certain health aspects. One such application is CalFit [5]. CalFit uses the native sensors on specific Android phone models to calculate energy expenditures based on the SPINE Kcal algorithm. It then logs the user's data into a database where all the users of this application can compare their caloric expenditures. The database also ranks users and allows for the creation of teams to sponsor encouragement and competition.

Work has also been done to use artificial neural networks, Bayesian networks, and Hidden Markov Models to develop context aware sensing in BSNs. People are very sensitive to external context changes such as a change in the person's activity or temperature of the environment and these situations need to be analyzed appropriately to draw the correct conclusions about a person's health status. Difficulties in accomplishing this include noise introduced by the sensors, the need for context sensing to detect transitions in context as opposed to a single snapshot in time, and the problem that as the number of inputs (sensors) increases, the learning rate slows down. Advanced computing techniques such as neural networks and Bayesian networks would be advantageous because each of the individual sensors could learn without supervision and do not require prior knowledge of the context [19].

## 4. SYSTEM OVERVIEW

To further the functionality and diversity of body sensor networks, we propose HatterHealthConnect (HHC). HatterHealthConnect is a health monitoring system that gathers physiological information to be integrated with social networks to promote healthy living and further peer connectivity. Live data such as heart rate, muscle activity, and workout duration are all calculated and sharable through these networking portals. HatterHealthConnect is designed to be easy to use, highly portable, and unobtrusive. Figure 1 depicts the overall design of HatterHealthConnect.

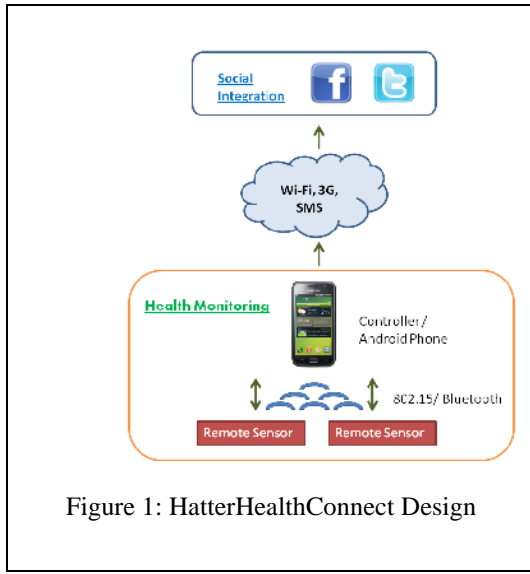

Figure 1: HatterHealthConnect Design

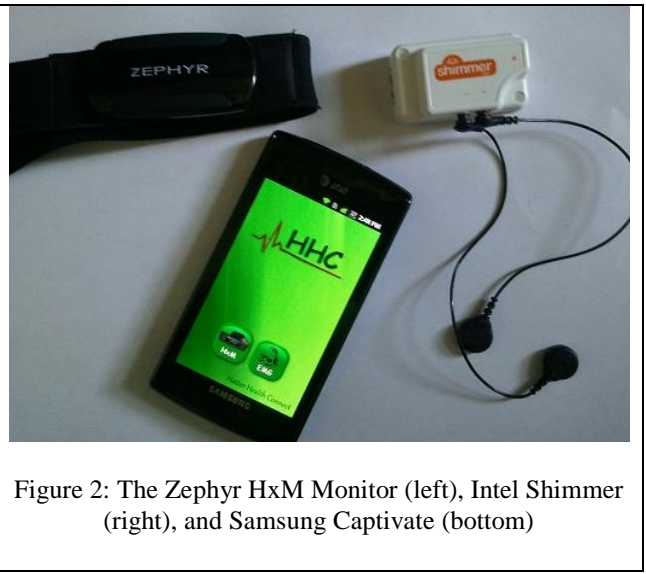

Figure 2: The Zephyr HxM Monitor (left), Intel Shimmer (right), and Samsung Captivate (bottom)

The monitoring and gathering of physiological data is handled by a wireless body sensor network comprised of multiple nodes. As shown in Figure 2, the sensor nodes include an Intel Shimmer Wireless Sensor Unit and attached EMG sensor and a HxM Bluetooth Heart Monitor. The coordinating or central node is a Samsung Captivate mobile device.

The BSN uses Bluetooth as its form of intra-node communication so the user can wear or carry the appropriate nodes wherever he/she goes. Once the user starts the application, the Samsung Captivate periodically collects data from the sensor nodes and store that information locally. Data collected on the phone is analyzed and processed further if necessary and the results can then be uploaded to the social networking sites Facebook and Twitter over Wifi or 3G.

This social interaction aims at promoting a healthier lifestyle by using encouragement and competition. Users are able to reach a much larger resource of advice and support as they share their own data and also comment on others' data. Also, by frequently uploading data, users are able to track their progress compared to their peers' and are positively pressured into continuing their own exercise activities or improving on them. Similarly, users are able to share their results via Twitter, so those following someone using HatterHealthConnect may be inspired to live a healthier lifestyle.

## 5. HARDWARE ARCHITECTURE

### 5.1 SHIMMER

Shimmer is a sensor platform designed for wireless applications that require the acquisition of data in real time. The Shimmer is designed to be low power and light weight in order to make wireless sensing efficient and unobtrusive in an everyday environment. The microprocessor and hardware components are designed to minimize power consumption for a long battery life and the total size of the Shimmer baseboard is only 53mm x 32mm x 15mm. The total weight of the unit is also only 22 grams, making the unit very versatile. Several key principles of the Shimmer wearable sensor are listed on their website and they are:

- Flexible: Each application can be customized to meet exact data capture and transfer requirements
- Configurable: Many different expansion modules/sensors are compatible with the Shimmer baseboard, including third party sensors. Shimmer can communicate over Bluetooth or 802.15.4 radio with any device that uses a similar radio.
- Open Source: The code is maintained and freely available at Sourceforge and GoogleCode
- Raw Data: Developers have full control over how the sensed data is handled and interpreted
- No Proprietary Software: Applications can be developed without having to match specific project output or interface requirements

The specific hardware components within the Shimmer were chosen and designed to meet the overall goal of making a low power and unobtrusive sensor. The MSP430 microprocessor was developed specifically to run on embedded, battery devices. The Shimmer makes use of almost every feature on the CPU, which controls all of the different peripherals attached. An integrated analog-to-digital converter captures sensor data, such as the attached EMG sensor. The Shimmer has the ability to store data internally via MicroSD flash storage, which allows data to be stored before sending or saved in case a network connection is dropped while the sensor is still gathering data [25].

Shimmer is an unobtrusive sensor platform because it communicates wirelessly. The Shimmer module contains two different radios: an 802.15.4 and a Bluetooth radio. IEEE 802.15.4 was developed specifically for low power personal area networks and is optimized to work within a short range. There are 27 total channels available with three different data rates available under this specification: 16 channels that have a data rate of 250 kb/s, 10 channels at 40 kb/s and 1 channel at 20 kb/s. 802.15.4 was specifically designed to be energy efficient at the physical and MAC layers, allowing it to have a very low power consumption level. The Shimmer contains a SR7 Radio module that communicates over 802.15.4 and has an indoor range of 5-10 meters.

Bluetooth is similar to 802.15.4 in that it is a short range, low power form of wireless communication. Bluetooth equipped devices can communicate at rates up to 3 Mbps. Bluetooth has 79 channels available that each have a data rate of 1 MHz. Unlike a network using the 802.15.4 protocol, which could contain up to 216 compatible devices, a Bluetooth network can only contain up to 8 devices. These individual groups of 8 devices can connect to form a larger group though, called a scatternet. The Shimmer device contains a Class 2 Bluetooth module that communicates through a 2.4 GHz antenna. This device has a range over 10 meters and can make use of all 79 channels. The overall goal of both protocols is to provide a short range, low power form of wireless communication. 802.15.4 has a slower data transfer rate, but it also provides more customization options and uses less power than Bluetooth. Bluetooth, on the other hand, is an older protocol and much more prevalent in mass market devices. Although 802.15.4 may better suit an application dealing specifically with body sensor networks, we have chosen Bluetooth due to its compatibility with Android devices [25].

## 5.2 EMG

Electromyography (EMG) measures the electrical impulses of the muscles within the body. It provides a quantifiable way to view the activity of a muscle while at rest and throughout the entire range of motion of a movement. The two different EMG techniques are intramuscular and surface EMG. Intramuscular EMG involves an electrode needle being inserted into the muscle tissue to target specific motor units. The muscle should behave in a certain manner in reaction to the needle, and results are compared to that. On the other hand, surface EMG (SEMG) is a noninvasive technique that allows EMG data to be measured without penetrating the skin and provides for a much broader evaluation of a muscle. SEMG picks up the electrical signals that are fired from a population of motor units within a muscle. These electrical signals travel through tissue until they eventually reach the surface of the skin, where electrodes then sense the energy. SEMG is a much more appropriate technique for sampling users in a workout environment and the data gathered will be more appropriate because it relates to a wider range of muscle use.

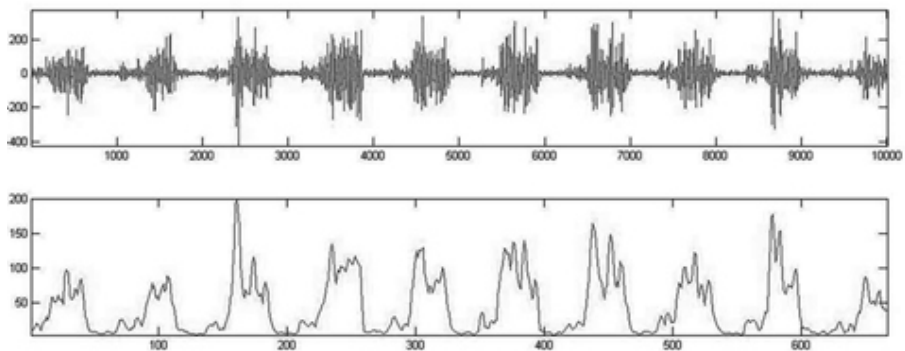
Figure 3: Raw EMG data (top) and its Root Mean Square (bottom) [37]

Once data is gathered from the electrodes, it must be processed before it becomes easily understandable. The initial type of processing is filtering, or removing the unnecessary data from the sample. This attempts to remove the electrical noise that is caused by anything other than the electrical impulses in the muscle. Once the data is properly filtered, the data can either be analyzed in its raw form or it can be further processed. Displaying the raw EMG data in a graph creates an oscillating line, which contains both positive and negative values as shown in Figure 3. From this graph, a user can see when the muscles are activated by looking at the thickness and height of the line. Although this may allow users to get a quick look at their muscle's energy expenditure, it may be harder to draw conclusions from the data without further processing it. To make the data easier to view, all processing techniques first rectify and smooth the data. Rectifying converts the negative electrical potential and adds it to the positive potential, making all values positive. Smoothing the data is done by integral averaging, which involves averaging a set of points and plotting that average for each point instead of plotting each individual point. This reduces variability in the data because now outlying values are averaged with their surrounding values to form a smoother graph [26].

Aside from processing raw EMG data for display, several processing techniques are also used to quantify the data, yielding numbers that can more easily describe muscle energy expenditure. Peak-to-peak measuring calculates the difference between the top and bottom of each trace and averages this value over a period of time. Integral averaging, the same method used for smoothing EMG graphs, can be calculated and represents .637 of one half of the peak-to-peak value. Root Mean Square (RMS) is a method that squares the data, calculates the average, and then calculates the square-root of this value. RMS is more commonly used than integral averaging because it provides less distortion [26]. This processed EMG data can now be used to evaluate criteria such as:

- The activation timing of a muscle: when energy expenditure of a muscle begins and ends and how frequently that occurs within a set period of time.
- The symmetrical expenditure of muscles: whether symmetrical muscles, such as the left and right bicep, display the same muscle activity levels for an exercise.
- A fatigue analysis on the muscle: how quickly or slowly the muscle decreases in energy expenditure [27]

The Shimmer EMG sensor is an attachment to the main Shimmer board and maintains the lightweight and small form factor of the unit. The EMG attachment is a surface EMG that connects to the skin via disposable electrodes and captures the activity of the entire muscle. The data gathered is filtered and the integral average is calculated before storing the data [28].

## 5.3 ZEPHYR HXM

The HxM connects with a mobile device over a Bluetooth link. It supports only one connection at a time and sends messages at a rate of one per second while connected as seen in Figure 4.

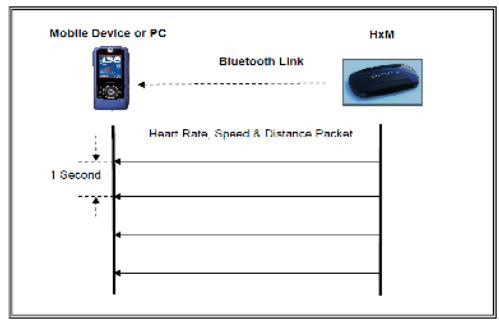

Figure 4: HxM Bluetooth Link

The HxM operates in simplex mode: it only sends data packets and does not process received packets. The discoverable name of the device is HXMxxxxxx where xxxxxx is the serial number of the sensor and its passkey is 1234. The HxM uses the Bluetooth SPP (Serial Port Profile) to connect to another device and uses the following structure:

International Journal of Computer Networks & Communications (IJCNC) Vol.5, No.2, March 2013- 115,200 baud
- 8 data bits
- 1 stop bit
- No parity [28]

The body sensors connecting over Bluetooth to HHC must rely on their own batteries for data transmission. Both the Zephyr HxM and Shimmer sensors are equipped with rechargeable batteries which can be recharged by connecting to their corresponding charging stations included with the products. The HxM has a standard battery life of 26 hours when activated due to its constant streaming of sensor information. The Shimmer sensor's battery life varies greatly depending on the frequency the data is being sent. Because this value can be adjusted, battery life is not constant after every use. It is recommended to recharge the sensors after every use to keep the devices prepared for next use.

## 5.4 BASE STATION

The hardware architecture for the base station, which is running the Android platform, is relatively up to the discretion of the manufacturers. Any new additional hardware added to the platform requires software support from the manufacturer as well. However, there are common pieces of hardware usually supported by the platform. These features include telephony (EDGE, 3G, Voice), local data connections (WiFi, Bluetooth), Camera, GPS, compass, and accelerometer. The Linux kernel is responsible for the driver connection to the hardware, and once implemented, developers can access the hardware through high-level application framework calls. While most phones that run the Android platform will work for HatterHealthConnect, this project specifically uses the Samsung Captivate.

The Samsung Captivate, or Captivate, is an AT&T variant of the Samsung Galaxy S and is a high-end Android smartphone with many of the latest hardware features [29]. Those features include:

- WGVA 4' Touch Screen
- Bluetooth Capability
- GPS
- Wi-Fi
- 5 MP Camera

Although most phones currently support Bluetooth, it is an important feature to include when discussing hardware architecture. Most low power consuming sensors use a wireless protocol for connection, and because the Captivate accepts only Bluetooth, it is important to find sensors which connect over Bluetooth.

## 6. SOFTWARE ARCHITECTURE

### 6.1 TINYOS

TinyOS, initially a project at UC Berkeley, is an open source operating system specifically designed for wireless sensor networks. This embedded operating system is completely non-blocking, so the majority of input and output functions are asynchronous and require a callback. TinyOS and applications that run on it are written in nesC [30]. The nesC (network embedded system C) language is a programming language similar to C, and designed to follow the execution model of TinyOS. NesC is made up of components, which encapsulate state and functionality similarly to objects in an object-oriented language. However, each of these components can only reference its local namespace. In order for one component to call a function from another component it must explicitly declare that function as well. In this way, function pointers are not needed and all connections are made during compile-time. This is possible because mainly because the embedded systems TinyOS and nesC are running on generally have specific tasks that do not require dynamic program loading due to user input. The behavior of each component is represented by a set of interfaces. Each interface either contains some functionality that is accessible to the user or represents some functionality the component needs to complete a task. Interfaces provide a way for all the individual components to be linked together to form a program. Overall, nesC was designed to provide a statically-linked environment that allows for greater runtime efficiency in the embedded system environment [31].

### 6.2 BIOMOBIUS

The EMG application run in TinyOS on the Shimmer node is a part of the BioMOBIUS research platform. This research platform was developed by TRIL Centre to allow the rapid creation of applications involving biomedical monitoring devices and sensors. BioMOBIUS is aimed at those who need to monitor the activity and physiological information of their users or research subjects. This platform supports many third party sensors and devices, including the Intel Shimmer. One of the main goals of the BioMOBIUS platform is to allow individual components to be encapsulated into individual "blocks." Once a single component is created, other users do not have to replicate it, they can simply reuse that component. The code behind the BioMOBIUS platform is also open and shared so that developers can make alterations to the source code to fit their specific needs [33].

## 6.3 COMMUNICATION PROTOCOLS

### 6.3.1 SHIMMER PROTOCOL

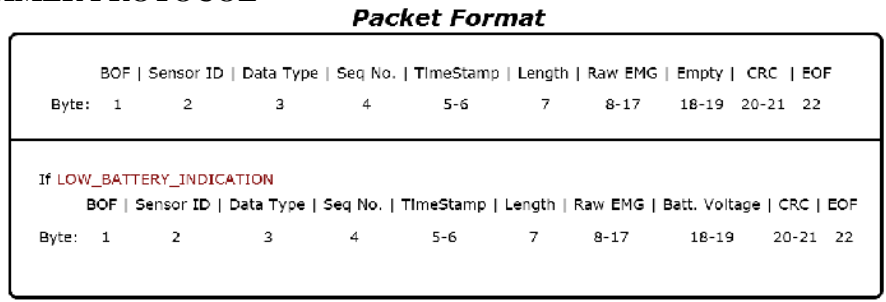

Figure 5: Shimmer Packet Format

The format of the data is identical in all instances except when the Shimmer has received a low battery indication from the TinyOS operating system. This low battery indication stops the streaming of data once the battery voltage drops below a regulator value of 3V. By default, the EMG sensor samples data at 500hz and each packet of data sent by the Intel Shimmer to the Android base station exactly contains fourteen bytes of data as shown in Figure 5. The MSP430 CPU is byte addressed and little endian, so slots of data that take more than one byte to hold the data, such as the time stamp, are sent with the highest addressed bytes followed by the lowest addressed bytes. In each packet, the first and last bytes are simply indicators of the beginning and end of each packet. The Sensor ID byte contains a unique identifier for each sensor, which is followed by a static byte used to describe the type of data being sent. This Data Type byte allows the base station to distinguish the difference between similar Shimmer packets that contain a different type of data, such as ECG data. An incrementing sequence number and timestamp are used to record when the data was taken and to make sure the packets come in order. The actual EMG data and its length are stored in bytes seven through nine. The tenth and eleventh bytes are used to display the battery voltage remaining if the low battery indication has gone off on the Shimmer, otherwise these bytes are empty. Finally CRC is calculated and stored so the data can be validated on the base station.

### 6.3.2 HXM PROTOCOL

The HxM follows a very simple structure when transmitting messages. As mentioned in the hardware architecture section, the HxM only sends data and does not process any received data. The basic message format consists of a few bytes for handling message processing along with the actual payload [29].

As shown in Figure 6, the message includes the following:

- **STX** - The start of text ASCII control character which signals the start of the message.
- **Msg ID** – uniquely identifies the HxM message and is in binary format. The standard data packet ID for the HxM is 0x26.

- **DLC** – Data Length Code specifies how many bytes of information are located within the payload between zero and 128 (inclusive).
- **Data Payload** – The actual data recorded by the HxM and can contain anywhere between zero and 128 bytes of data.
- **CRC** – An 8-bit CRC
- **ETX** – The end of text ASCII control character which signals the end of the message.

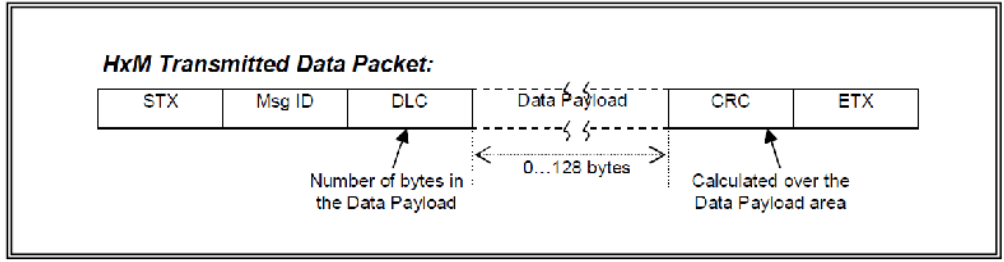

Figure 6: HxM Packet Format

The Data Payload contains the standard data message from the HxM. The message contains the sensor data such as heart rate, speed, and distance. The past 14 heart rate measurements timestamps are sent as well so packet loss can be determined. This is useful when calculating the average values from the biological data. The standard data message can be seen in Figure 7.

Figure 7: HxM Standard Data Message

The payload of the data message contains more information about the sensor itself such as Firmware ID, Hardware ID, and Battery Charge Indicator, along with the actual sensed data. The remaining data is structured as follows [29].

- Heart Rate – an unsigned byte with a valid range between 30 and 240 bpm. If no heart rate is detected, 0 is reported.

International Journal of Computer Networks & Communications (IJCNC) Vol.5, No.2, March 2013

- Heart Beat Number – unsigned byte that is incremented each time a heart rate is detected. Rolls over at 255.
- Heart Beat Timestamps (1-15) – 16 bit unsigned integer representing the heart beat timestamp between 0 and 65535 milliseconds. Rolls over at 65535.
- Distance – 16 bit unsigned integer representing the total distance travelled in 16ths of a meter. Rolls over ever 256 meters.
- Instantaneous Speed – 16 bit unsigned integer representing the instantaneous speed of the device in steps of 1/256m/s. The valid range is between 0 and 15.996 m/s.
- Strides – unsigned byte representing the number of strides the wearer has taken between 0 and 255. Rolls over at 255.

### 6.4 ANDROID OPERATING SYSTEM

The Android is a mobile platform based on a software stack rich with features. It includes an operating system, middleware, and key applications which can each be adjusted individually to maximized platform performance. Some of the key software features Android includes are the Dalvik virtual machine, integrated browser, optimized graphics, SQLite, and media support [34]. The Android architecture is essentially divided into four layers as can be seen in Figure 8. The layers are built on top of each other, the lower layers vital for upper layer development. While the majority of development of HHC was focused on the Application layers, this section briefly discusses the function of the individual layers.

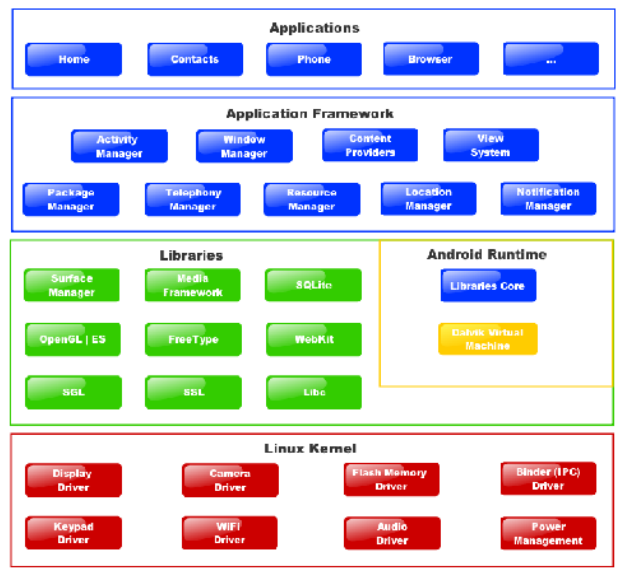

Figure 8: Android Operating System

**Linux Kernel**

The current Android Linux kernel is based on Linux 2.6 for its core services. It handles low level system services such as security, memory management, process management, network stack, and drivers. With the drivers, it is responsible for acting as an abstraction between the hardware and other layers in the stack [34]. As seen in the hardware section, customization to the kernel can support different hardware technologies included by manufacturers.

**Android Runtime**

As applications for Android are primarily written in Java, it includes a set of core libraries to provide function for the Java programming language. Each application runs on an instance of the Dalvik virtual machine, which has been optimized for mobile development. It has a minimal memory footprint and relies on the Linux kernel for low level functionality such as memory management, threading and networking [34]. In Android version Froyo (2.2) and beyond, the Dalvik virtual machine includes a JIT (Just in time) compiler to greater improve the speed of running code. Code is actively analyzed while running and translated into a faster form all while using little memory. The JIT is just one of the enhancements made to the Android runtime to make it highly efficient on mobile devices [35].

**Library**

In addition to the Java libraries included in the runtime, Android also includes a set of C/C++ libraries which can be accessed through the Application framework. The libraries handle lower operations such as media, 2D graphics, 3D graphics, SQLite, and bitmap and vector font rendering [34]. The Native Development Kit (NDK) provides direct access to several of these libraries and can be used to create classes that may have a slight improvement over their Java counterpart [36].

**Application Framework**

Android allows developers to deeply integrate their applications with the operating system because of its open platform. Applications have access to the same framework APIs all the core programs have and can take advantage of any currently designed components. The framework does have security constraints, but if permissions are granted by the user, applications can publish their capabilities to any other applications. Below all applications is the core of the framework which includes Views, Content Providers, a Resource Manager, a Notification Manager, and an Activity Manager [34].

**Application**

The application layer on the Android platform is mostly left to third party developers. Developers can use the other layers, specifically the application framework, to develop applications in Java. Most stock versions of Android come with default applications such as Phone, Browser, and Contacts and developers can use them as examples for other applications. With the Software

Development Kit (SDK) for Android developers can create deeply integrated and graphically pleasing applications for users.

## 6.5 SERVER SOFTWARE

The server-side of HatterHealthConnect is written in PHP. The first objective of this server is collecting, storing, and summarizing the data uploaded from individual Android base stations. MySQL is the chosen relational database management system for the server side of HHC. All necessary information for HHC is stored within this database, including biophysical, user, and team data. This database is built to support the individual Android phones and also the Facebook and Twitter applications.

The second part of HatterHealthConnect's back-end server is to provide an interface where users can view data and interact with other users. This is accomplished using social networking sites. These social networking sites often provide developer-friendly APIs for accessing and updating user data. Facebook and Twitter support APIs for a variety of platforms and languages. These APIs often include multiple ways of accessing the site data and have security measures in place to restrict whose data is visible. Additionally, many open-source projects have been initiated to wrap these APIs and increase the ease of site interaction. Facebook development is more complex than Twitter because of all the different functionality Facebook provides. User's status updates are only a small portion of the site. These applications can access third party information and can use pre-made tools by Facebook for enhancement.

Facebook applications are web applications that are built in a common web programming language, such as PHP, and then loaded into the context of Facebook. Essentially, a web application is designed separately from Facebook on a server and then is linked from within Facebook when the app is run. The benefit of using Facebook comes from all the information that Facebook provides the application about the user. Once a user agrees to share their Facebook information with the application, the application is able to make Facebook API calls to access his or her user id and then link the id to the proper data in the database. The web application for HatterHealthConnect is written in PHP and located on the same server as Android-facing scripts.

## 7. IMPLEMENTATION DETAILS

### 7.1 BASE STATION IMPLEMENTATION

The Android portion of HHC is designed similarly to the majority of Android applications. The following sections describe the class structure of HHC along with the design implementation

### 7.1.1 BASE STATION CLASS DESIGN

The source code for HHC is structured such that Java class files are separated from the XML files. The XML defines layout, styles, and global strings which can then be referenced from the Java code.

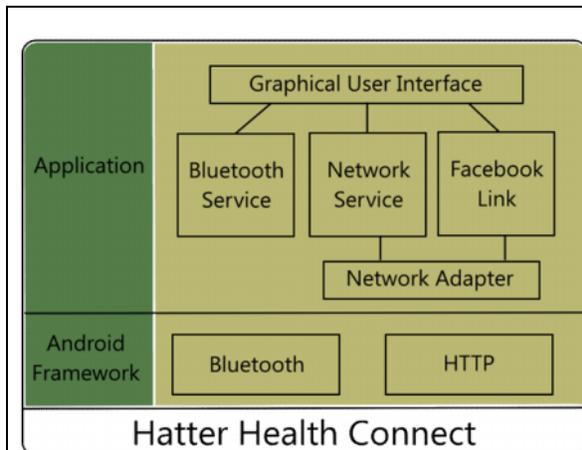

Figure 9: HatterHealthConnect program structure

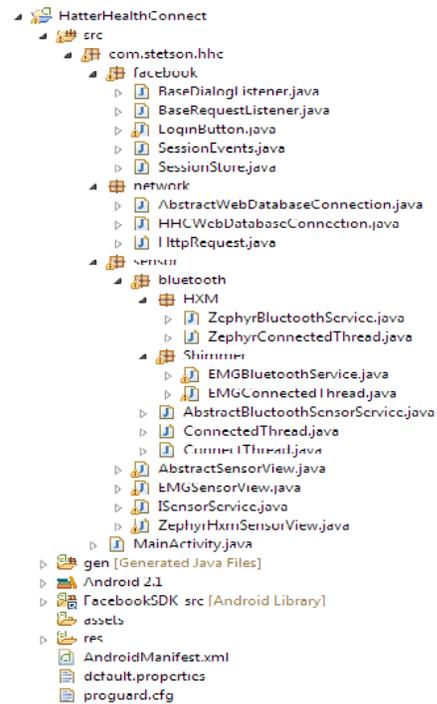

Figure 10: HHC Package Structure

As illustrated in Figure 9, the lower levels of HHC interact directly with the Android framework to share data with other devices. Bluetooth calls are made to body sensors through the Bluetooth service to handle remote data collection. The data is then processed and passed either directly to Facebook through the Facebook Link classes, or uploaded to the backend database over HTTP. User interaction is handled through the GUI. The user is able to choose how his/her data is uploaded and at what frequency. The program can also be navigated through the menu to change sensors and handle reconnections.

Java development allows class files to be separated into packages to increase encapsulation during development. The package structure of HHC can be seen in Figure 10. Each piece of the application is separated out into one of the following packages or classes:

- **com.stetson.hhc.facebook** – Handles Facebook authentication along with any buttons or views needed to connect. When authenticating to Facebook, it checks to see if an instance of the Facebook application is currently installed on the device. If so, it authenticates through the application, otherwise opens up a web view to authenticate.
- **com.stetson.hhc.network** – Handles HTTP and remote database connections over the Internet. Includes an abstract web database connection class for extensions to other databases.

- **com.stetson.hhc.sensor** – Contains the majority of the code handling sensor connections along with class creating the GUI.
- **AbstractSensorView** – Abstract class serving as the base view for any GUI needing to connect to a sensor. Contains methods for connecting and handling sensor information.
- **EMGSensorView and ZephyrSensorView** – AbstractSensorViews each dedicated to a specific HHC sensor.
- **com.stetson.hhc.sensor.bluetooth** – Handles Bluetooth sensors such as the HxM and Shimmer EMG. Base packages *com.stetson.sensor.bluetooth.HXM* and *com.stetson.sensor.bluetooth.Shimmer* extend AbstractBluetoothService and ConnectedThread to create a fully functional sensor connection.
- **AbstractBluetoothSensorService** – Establishs an initial connection to a sensor and manages the connection by sending messages to the extending class.
- **ConnectThread and ConnectedThread** – Used by a sensor service to connect to a sensor and handle any information sent to the device from the sensor.

### 7.1.2 BASE STATION DESIGN

The HHC Android application was designed with both functionality and user perspective in mind. The application needed to contain the necessary pieces to perform a variety of useful functions, while at the same time feeling very intuitive. The resulting application ended up with a very simple design, with much of the connection work being done in the background, and the user having the options to handle where the data is uploaded.

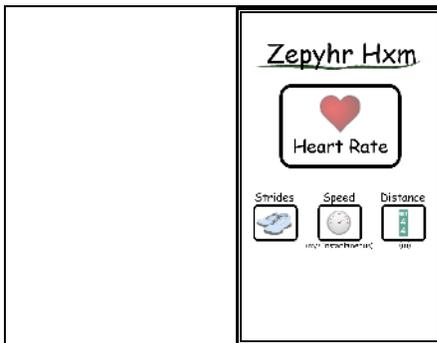

Figure 11: Screenshot of HHC heart monitor view

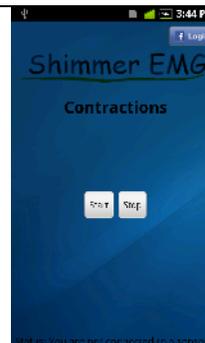

Figure 12: Screenshot of HHC EMG monitor view

In order for the user to start recording data, they simply select which attachment he or she is using and then begin the connection. The application then connects to the sensor and data is passed back to the phone. Within each of the individual views a user can also log into Facebook, or directly post current information onto his or her Facebook wall. Figures 11 and 12 show snapshots of the Android Application for heart monitor and EMG monitor respectively

Security is an important part of an application when dealing with personal information such as health data. HHC maintains security by only submitting/posting data the user has willfully decided to submit. The user has full control over what data is uploaded to the backend or social networking sites. The base station does not transmit any data without first being prompted by the user.

## 7.2 SERVER IMPLEMENTATION

### 7.2.1 USER DATABASE

The BSN-facing side of the server is a collection of PHP scripts that wait for data to be sent via POST to the server's main receiving script. These PHP scripts evaluate the type of data being sent and store it appropriately into HatterHealthConnect's database. As shown in Figure 13, this database is made up of several tables that can easily be queried for future data retrieval. The table "Users" has been created to store data that allows HHC to link users to their appropriate Facebook and Twitter accounts. A "Teams" table contains the data pertaining to the many different groups created through the Facebook application, including their unique name. Individual users can be a part of several teams and each team can have many users so a "UsersToTeams" table is defined to map those relationships. Finally, the "Workouts" table contains all the necessary information about every individual workout the user performs. Information such as the day, time, and duration are included. The data passed from the health sensors is stored in separate tables specific to the sensor gathering the data and then linked to a workout. This way, more sensors could easily be added by creating more sensor-specific tables and linking them to a specific workout. The social networking applications can now pull data about the frequency, length, and health data of workouts by querying this database.

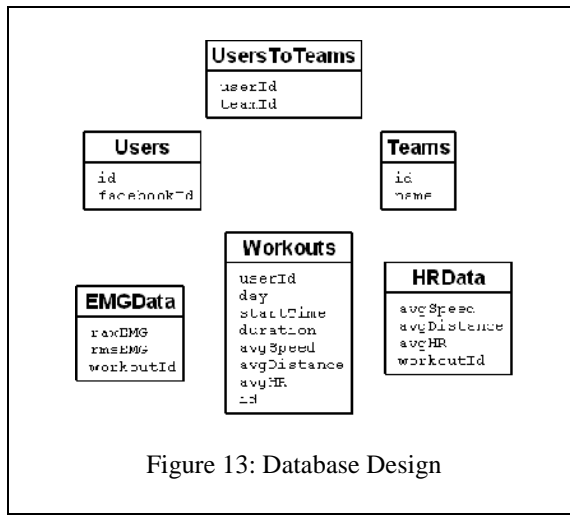

Figure 13: Database Design

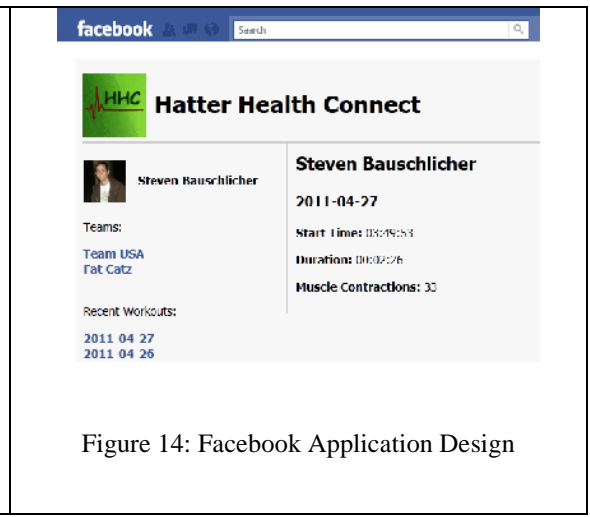

Figure 14: Facebook Application Design

### 7.2.2 SERVER APPLICATION DESIGN

A user has several ways of sharing or storing the physiological and other workout data once he or she has finished a workout. HatterHealthConnect includes a Facebook application as the primary method of exposing content to users. Figure 14 shows Facebook application for EMG monitor. The Facebook application pulls data from a database that contains uploaded data from the users and displays it to them. Users are able to see their progress as well as others connected with them.

By being able to view their own workout history, users are able to see if they have improved or view the results of different training patterns. Within the application users can create or join groups so that data from other members that belong to the same group can be accessed and compared against a user's own health data. There are also several other features that make use of Facebook that do not require a user to access the web application. For instance, Facebook has released an open source API for Android, allowing a user to interact directly with his/her Facebook account. The Android application, if prompted, can then update a user's status based on the information received from the healthcare monitoring portion. Interacting directly through the Facebook Android API for status updates should increase awareness, because it will require little work to update one's status and many users should use it. It will be much easier for users to activate the updates on their phones than to go through the Facebook web application and update it there.

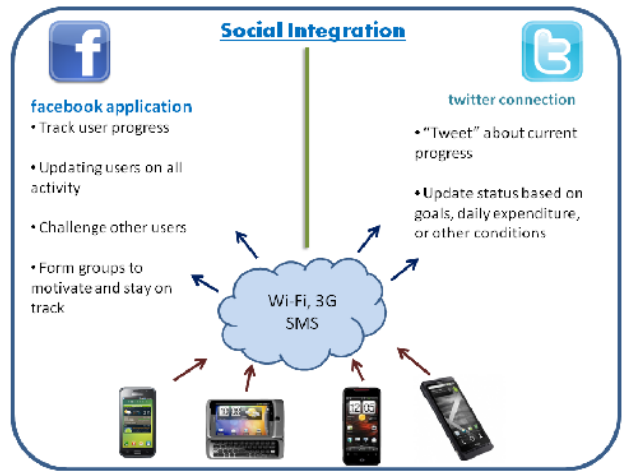

Figure 15: Methods of Social Interaction

Twitter integration is similar to the status update portion of the Facebook integration. Twitter is a micro blogging site that lets users "tweet" updates about themselves, and remotely interacting with the site is fairly simple. A HatterHealthConnect user has the option to tweet about his/her progress or current status, very similar to how one would update his/her Facebook status. Along with Facebook, Twitter should provide another channel for promoting health and fitness by creating exercise awareness for peers. Figure 15 shows the methods of different social interaction.

## 8. CONCLUSION

Body sensor networks continue to play an important role in the development of healthcare applications as the need for lightweight and remote monitoring continues to grow. In this paper we have presented HatterHealthConnect, a network that combines the mobility and monitoring of a BSN with the interaction capabilities of social networks. With the advent of social networking and the increased ability to share data, people are able to communicate like never before. We believe that creating a tool to connect health/sensor data to a social media channel will help promote physical wellness by creating peer groups that can help motivate and encourage each other. HatterHealthConnect includes a reliable BSN that is easy to operate, as it can run on any mobile phone the uses Android OS 2.1 or greater and is completely wireless. Promoting physical wellness is an important goal to focus on and self motivation is not always enough to encourage it. HatterHealthConnect is designed to meet that goal effectively by easily connecting users to a global network of those who want to pursue a healthier lifestyle. Nowadays more and more people are comfortable using social networks and will not have any difficulty to access and interact with data on these sites. Users can choose to share information with a group of friends or only with their doctors. They can also choose not to share any data, but use the system as a method to keep track of their exercise routine and monitor their own progress. Our preliminary evaluation of the survey completed by twenty three users with different demographic data such as age, social network usage, and workout rate show that it is reasonable to assume users who already workout will be quicker to adopt HHC and without too much concern. However, older users who do not use social networking sites may be slower to adopt the application.  We think that this application can be more popular among youth. We hope that it could act as a tool to promote health awareness that could yield in solving child obesity problems.